# Development of COTS ADC SEE Test System for the ATLAS LAr Calorimeter Upgrade[*]

HU Xue-Ye(胡雪野)[1,2]   CHEN Hu-Cheng(陈虎成)[3]   CHEN Kai(陈凯)[3]   Joseph Mead[3]   LIU Shu-Bin(刘树彬)[1,2]   AN Qi(安琪)[1,2]

[1]State Key Laboratory of Particle Detection and Electronics, University of Science and Technology of China, Hefei 230026, China

[2]Department of Modern Physics, University of Science and Technology of China, Hefei 230026, China

[3] Brookhaven National Laboratory, Department of Physics, Upton, New York 11973, United States

**Abstract:** Radiation-tolerant, high speed, high density and low power commercial off-the-shelf (COTS) analog-to-digital converters (ADCs) are planned to be used in the upgrade to the Liquid Argon (LAr) calorimeter front end (FE) trigger readout electronics. Total ionization dose (TID) and single event effect (SEE) are two important radiation effects which need to be characterized on COTS ADCs. In our initial TID test, Texas Instruments (TI) ADS5272 was identified to be the top performer after screening a total 17 COTS ADCs from different manufacturers with dynamic range and sampling rate meeting the requirements of the FE electronics. Another interesting feature of ADS5272 is its 6.5 clock cycles latency, which is the shortest among the 17 candidates. Based on the TID performance, we have designed a SEE evaluation system for ADS5272, which allows us to further assess its radiation tolerance. In this paper, we present a detailed design of ADS5272 SEE evaluation system and show the effectiveness of this system while evaluating ADS5272 SEE characteristics in multiple irradiation tests. According to TID and SEE test results, ADS5272 was chosen to be implemented in the full-size LAr Trigger Digitizer Board (LTDB) demonstrator, which will be installed on ATLAS calorimeter during the 2014 Long Shutdown 1 (LS1).

[*]This work was supported in part by the Unites States Department of Energy Contract No.DE-AC02-98CH10886.

# 1 Introduction

The ATLAS LAr Calorimeter upgrade project is proposed to enhance the physics reach of the experiment in the high-luminosity environment foreseen in the next 10 years [1]. In order to provide higher-granularity, higher-resolution and longitudinal shower information from the calorimeter to level-1 trigger processors in this upgrade, new LAr calorimeter trigger readout electronics need to be designed, built and installed. Compared to the existing readout cell "trigger tower", the new readout element called Super Cell [1], which is a 10-fold finer granularity scheme, also provides additional information and more powerful tools to the Level-1 trigger feature extraction. Also, the digitization precision of the Super Cell signals is improved by at least a factor of 4 compared to the existing Level-1 system by optimizing the quantization scale and the dynamic range of the digitizers. These upgrades will be essential to extend the physics potential at higher instantaneous luminosities and the more severe pileup conditions expected after Phase-I and Phase-II upgrades of the LHC.

In the Phase-I upgrade, ~40,000 channels of super cell signals will be digitized at the front end LTDB, and data will be streamed out to the back end DPS (digital processing system). A radiation tolerant ADC is required for signal digitization in the front end electronics. The LAr collaboration has prepared two different technological routes: custom ASIC ADC development and COTS ADC evaluation. Two custom ASIC ADCs are under different stages of prototyping developments. However, given the uncertainty in the development cycle and costs associated with a custom chip design, an extensive study has been conducted for a COTS option that meets both electrical and radiation requirements. Previous studies on the radiation sensitivity to many COTS parts can inform component decisions appropriate for our design. Good experience can be found in Reference [2]-[5].

In this paper, we present ADS5272 [6], a TI COTS ADC as a good candidate for use in the LAr calorimeter electronics upgrade. The electrical features of the ADS5272 include: 65MSPS maximum sampling rate, 12 bit dynamic range, 11.5 resolution (ENOB), 162.5ns (6.5 clock cycles) latency, 113mW power consumption per channel. These parameters meet the digitization requirements [1] of the LAr calorimeter upgrade, and therefore a test program was developed to study the ADS5272 radiation characteristics. The outline of this paper is organized as follows. In Section 2, we discuss the TID radiation effects and their tolerance to an ionizing dose of 17 COTS ADCs. In Section 3, we show the detailed development of the ADS5272 SEE evaluation system, consisting of hardware preparation, firmware development and software application. In section 4, we use our evaluation system to characterize the ADS5272 SEE radiation tolerance in irradiation tests. In Section 5, we conclude this paper, and summarize what we achieved.

# 2 Total Ionization Dose (TID) Irradiation Test

Long-term exposure to ionizing radiation can cause parametric degradation and ultimately functional failure in electronic devices. The damage occurs via electron-hole pair production, transport and trapping in the dielectric and interface regions.

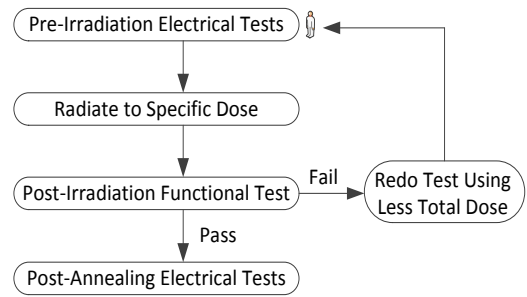

Fig. 1. Diagram of TID test flow

| COTS ADC | Dynamic Range [bit] | Max Sampling Frequency [MSPS] | Analog Input Span [V p-p] | Number of Channels per Chip | $P_{total}$ per Channel [mW] | Technology | Vendor | TID [kRad(Si)] |
|---|---|---|---|---|---|---|---|---|
| AD9265-80 | 16 | 80 | 2 | 1 | 210 | 0.18um CMOS | ADI | ~220 |
| AD9268-80 | 16 | 80 | 2 | 2 | 190 | 0.18um CMOS | ADI | ~160 |
| AD9269-40 | 16 | 40 | 2 | 2 | 61 | 0.18um CMOS | ADI | ~120 |
| AD9650-65 | 16 | 65 | 2.7 | 2 | 175 | 0.18um CMOS | ADI | ~170 |
| AD9253-125 | 14 | 125 | 2 | 4 | 110 | 0.18um CMOS | ADI | ~105 |
| LTC2204 | 16 | 40 | 2.25 | 1 | 480 | 0.18um CMOS | Linear | ~180 |
| LTC2173-14 | 14 | 80 | 2 | 4 | 94 | 0.18um CMOS | Linear | ~105 |
| LTC213 | 16 | 80 | 2 | 2 | 125 | 0.18um CMOS | Linear | ~100 |
| ADS4245 | 14 | 125 | 2 | 2 | 140 | 0.18um CMOS | TI | ~235 |
| ADS6445 | 14 | 125 | 2 | 4 | 320 | 0.18um CMOS | TI | ~210 |
| ADS5282 | 12 | 65 | 2 | 8 | 77 | 0.18um CMOS | TI | ~460 |
| ADS5263 | 16 | 100 | 4 | 4 | 280 | 0.18um CMOS | TI | ~2100 |
| ADS5294 | 14 | 80 | 2 | 8 | 77 | 0.18um CMOS | TI | ~1070 |
| ADS5292 | 12 | 80 | 2 | 8 | 66 | 0.18um CMOS | TI | ~1060 |
| ADS5272 | 12 | 65 | 2.03 | 8 | 125 | 0.18um CMOS | TI | ~8800 |
| HMCAD1520 | 14 | 105 | 2 | 4 | 133 | 0.18um CMOS | Hittite | ~2300 |
| HMCAD1102 | 12 | 80 | 2 | 8 | 59 | 0.18um CMOS | Hittite | ~1730 |

Table. 1. Diagram TID test results of COTS ADCs by June, 2012

To examine the effect of this issue on our COTS ADC, we performed TID tests with a Co-60 solid state gamma irradiation facility at Brookhaven National Laboratory (BNL). We have followed the test flow shown in Fig.1 and the results are shown in Table 1 [7].

Of the six ADCs which withstood doses larger than 1MRad (Si) (showed in the bottom six rows in Table 1), the ADS5272 is the top performer - surviving 8.8 MRad (Si), see Fig.2

Two ADS5272 samples have also been annealed by operating at ~85 degree after more than 2MRad (Si) TID test. After annealing, both ADCs recover to their original characteristics. Fig.3 shows the analog and digital power consumption of sample 2 before and after annealing.

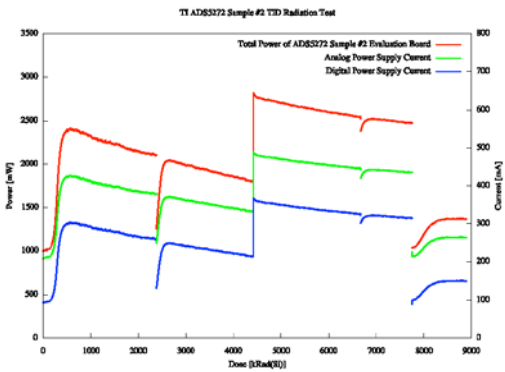

Fig. 2. Power consumption of ADS5272 #2 during the ~8.8MRad (*Si*) TID test

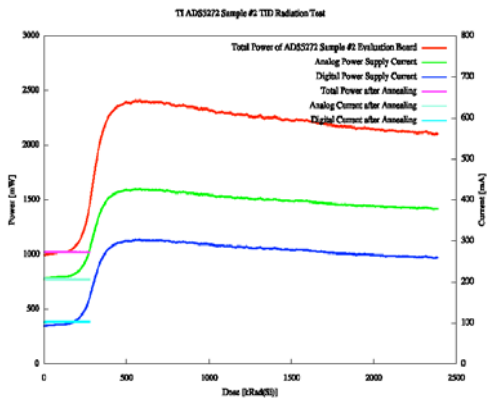

Fig. 3. Power consumption of ADS5272 #2 before and after annealing

## 3  Single event radiation effects

Since the TID irradiation test results of ADS5272 are very promising, we decide to set up an evaluation system for ADS5272 to characterize its single event effects. This system consists of three parts: 1) Hardware - an ADS5272 test board is custom built for SEE test, shown in Fig.4. 2) Firmware - we implement firmware in Virtex-6 FPGA on ML605 [8]-[9], which is generally responsible for acquiring data from the ADC, and controlling and monitoring it. 3) Software - it is developed in MATLAB GUI, which takes care of communication with ML605 through an Ethernet connection, sending configuration information and saving data for analysis in the case of a SEE.

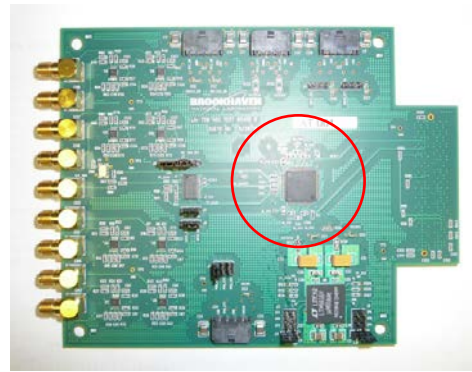

Fig. 4. A picture of ADS5272 test board

### 3.1 Design of ADS5272 test board

We have chosen SMA connectors as the input connectors for the ADS5272 test board. The output connector is a Samtec FMC (FPGA Mezzanine Card) HPC connector. This makes the ADS5272 test board easy to attach to the ML605. In order to make the input signals match with the ADS5272 differential full-scale input voltage range, we add an ADC driver in the analog signal chain. The ADC driver is ADI AD8138, which has been qualified up to 500krad TID. The AD8138 output and ADS5272 input are DC coupled with RC low pass filtering to improve the signal-to-noise ratio (SNR). Fig.5 shows the simulation of AD8138 circuit.

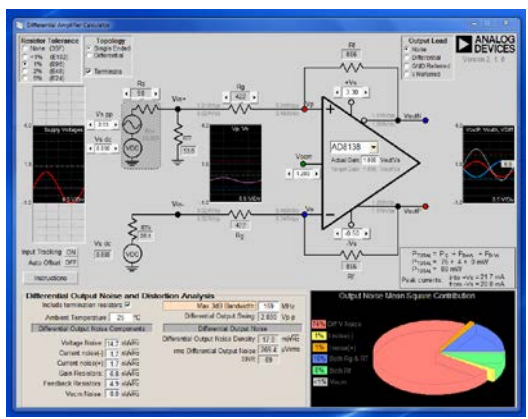

Fig. 5. Simulation of AD8138 circuit

The clock scheme for the ADS5272 has three options: SMA input clock, FPGA differential output clock and on board oscillator. ADS5272 needs a LVTTL clock, so we have chosen a Maxim MAX9160 to be a clock fan out driver for all clock input options, which has survived ~15MRad TID.

There are two options for the power supply: an external supply and an on board POL (Point-of-Load) DC-DC converter LTM4616 from Linear Technology. The external power supply is responsible for AD8138 power $\pm 3.3V$, ADS5272 analog power +3.3V and digital power +3.3V (contingency), ADS5272 reference voltage REFT +1.95V and REFB +0.95V, ADS5272 common-mode voltage +1.45V, and LTM4616 input +5V. The DC-DC converter LTM4616 can also provide analog and digital power to the ADS5272. A detailed block diagram of the ADS5272 test board is depicted in Fig. 6. It is worth mentioning that we keep a clearance red circle with 3 inches diameter (Fig. 4) around ADS5272 for the requirement of the irradiation test. No other active component is placed in this circle.

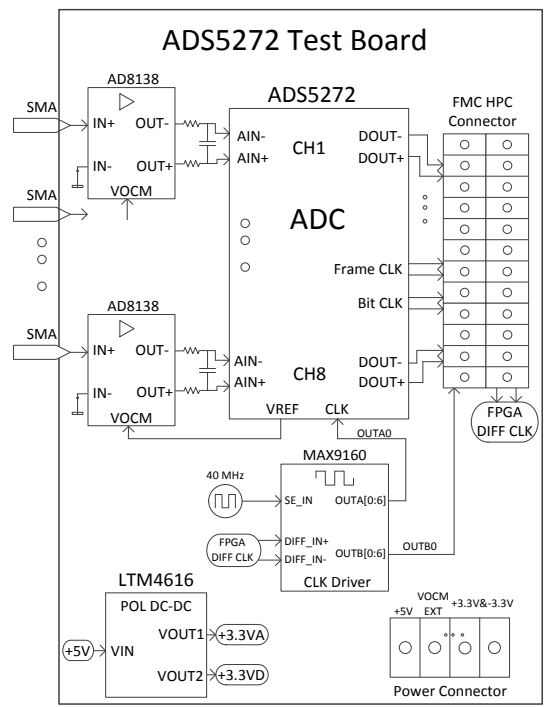

Fig. 6. Diagram of hardware development of SEE test system

### 3.2 Development of firmware in Virtex-6

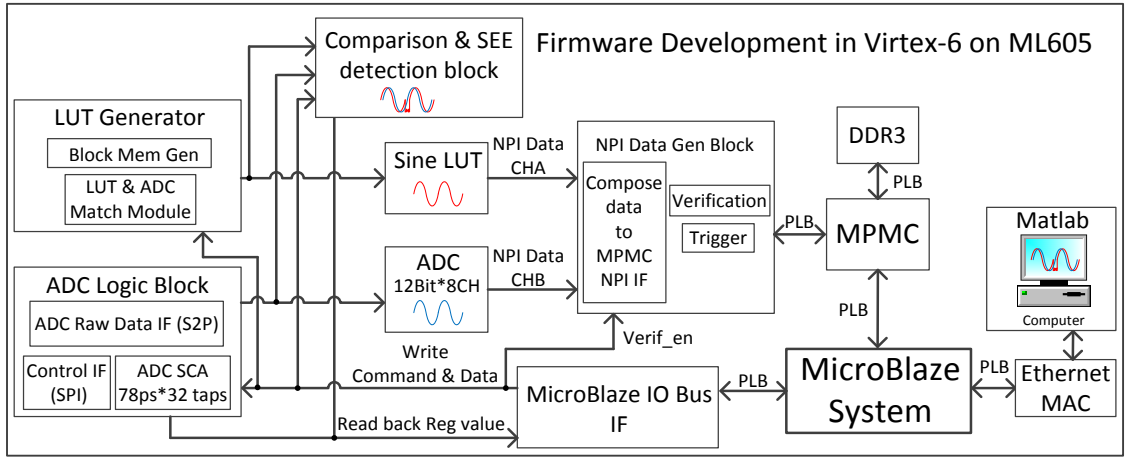

Fig. 7. Diagram of firmware development of SEE test system

The firmware of the ADS5272 SEE test system is developed in a Virtex-6 FPGA, which mainly consists of a MicroBlaze (UBLZ) core system and the FPGA fabric logic. The block diagram is illustrated in Fig.7. UBLZ is a 32 bit RISC (Reduced Instruction Set Computer) embedded processor soft core, which is generated to acquire data from the ADC and compare it with a LUT (Look-Up Table), then buffer the data to DDR3 SDRAM which can be read out through Gigabit Ethernet or USB. The FPGA fabric logic is comprised of five sub-function blocks: ADC logic block, LUT generator block, comparison & SEE detection block, NPI (Native Port Interface) data generator block and UBLZ IO bus interface (IF) block.

The purpose of ADC logic block is to de-serialize ADC raw serial data from serial to parallel (S2P), control and program ADC settings through a serial peripheral interface (SPI) and keep ADC bit clock 90° out-of-phase with respect to the data and frame clock through ADC sampling clock alignment (SCA). SCA function is realized through the adjustment of IODELAYE1 primitive of the FPGA according to associated SNR and noise floor plots.

The LUT generator basically aims to generate a programmable look up table via FPGA embedded block RAM resources. This look up table will be aligned and locked with the ADC waveform before the beam test starts. It is a critical preparation for comparison & SEE detection block.

Real time comparison of ADC data vs. LUT is done in the comparison & SEE detection block. This block can continuously check the difference between the ADC and LUT. When the difference is larger than the preset threshold which is programmable, we will deem the case as an SEE event. An error flag will then be polled to initiate DDR3 transfer and an error counter will start counting.

The NPI data generator block provides logic to compose ADC & LUT data according to PLB (IBM CoreConnect® Tookit Processor Local Bus) timing and data structure rules. The NPI block then sends data to the MPMC (Multi-Port Memory Controller) NPI interface. It also has a verification function to send user test patterns which is permitted by an enable signal (Verif_en) generated from UBLZ IO bus IF block.

The UBLZ IO bus IF block provides a read-write register interface, which is connected as a bridge to a UBLZ processor. It just leverages a simple user logic bus to decode a bunch of transactions. Write operations include transporting configuration information and delay tap values from MATLAB to ADC logic block, and updating the LUT according to the mean value of 100 samples calculated by MATLAB. It also sets the threshold to detect an SEE event and triggers the DDR3 test transfer as well. Read operations consist of getting values of the ADC registers and SEE error counter, examining SEE error flags and reviewing test patterns.

### 3.3 Software realized in MATLAB GUIDE

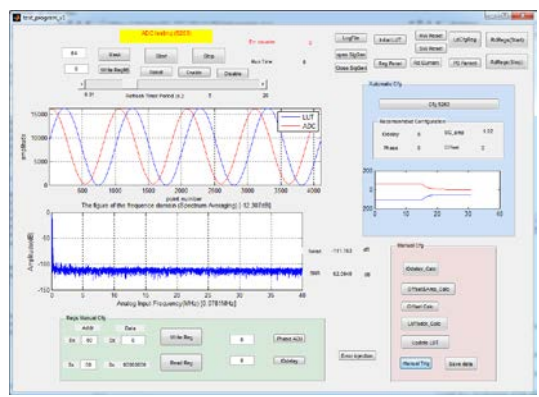

Fig. 8. A picture of SEE software GUI panel

Software is built on the GUI panel (showed in Fig.8.) with MATLAB GUIDE, which talks to the UBLZ system via TCP/IP server sockets.

MATLAB applications serve three principal functions. The first function is calculation and analysis. MATLAB calculates the mean value and RMS of the difference between ADC and LUT with 100 samples, sends the mean value back to the firmware "LUT Generator" block to update and match the LUT with ADC data. MATLAB also performs FFT with ADC data to get the corresponding noise level and SNR (Signal Noise Ratio) plot. It will then round the mean value of all working delay taps to get the most appropriate value for ADC SCA.

The second function is a control "keyboard". All input control information is manipulated by MATLAB. This includes issuing ADC hardware reset, sending ADC SPI configuration bits, adjusting the LUT address value and offset value, triggering a DDR3 transfer, setting the SEE error threshold, manually injecting an error to SEE test system for simulation and so on.

The third function is remote "monitor". MATLAB can plot the ADC data vs. LUT in real time and display ADC register values on the GUI. It can also monitor the voltage of the power supply and the amplitude of signal generator through Ethernet. This service is very convenient for debugging and testing.

## 4 SEE Test Results

The ADS5272 SEE evaluation system has been successfully used in multiple irradiation tests. In October 2012, an initial neutron beam test was performed at LANSCE WNR (Los Alamos, NM) with the maximum energy of about 800MeV. The neutron spectrum here is matched to that expected at the position of ATLAS LAr electronics crate. The second test was done in IUCF (Bloomington, IN), which will be presented in Part 4.2, to illustrate the correctness and practicality of the ADS5272 SEE test system. Another test is conducted in Mass General Hospital (Boston, MA) with 216MeV protons. The total SEU cross section observed in these tests are consistent with each other.

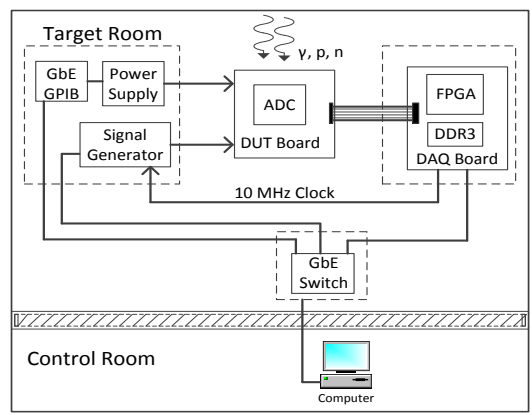

Fig. 9. Diagram of SEE test setup

### 4.1 Test setup

The beam tests described above share very similar test set up. We utilize a signal generator to inject a sine wave into ADS5272, which is running at $f_{sample} = 40MSPS$. The frequency of the sine wave is about $40kHz$ ($f_{sample}/$

$2^{10}$) to ensure enough samples are acquired for each cycle. The FPGA acquires ADC data and compares samples with the LUT in real time. Any deviation larger than a preset threshold will be flagged as an SEE event, and a record of ~4k samples is saved for posterior analysis. The system is synchronized by a 10MHz clock, which is generated by the ML605 board. The test setup diagram is shown in Fig.9 [7].

### 4.2 Proton Beam Test at IUCF

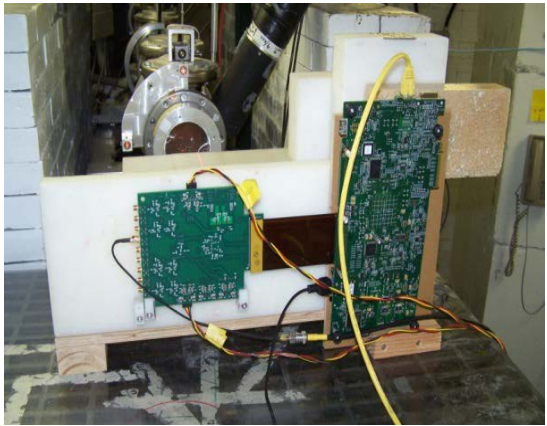

Fig. 10. ADS5272 SEE test at IUCF

One SEE test was performed at IUCF with high-flux proton beams on November 30th, 2012. A photograph of the IUCF test is shown in Fig. 10. A total of three ADCs were irradiated with ~200MeV protons to measure both SEU (Single Event Upset) and SEFI (Single Event Function Interrupt) cross sections.

| Sample No. | 1 | 2 | 3 |
|---|---|---|---|
| Fluence[$\times 10^{12} p/cm^2$] | 5.01 | 4.7 | 3.02 |
| No. of SEE Events (SEFI+SEU) | 17 | 19 | 3 |
| No. of SEFI Events | 4 | 6 | 1 |
| No. of SEU Events (with Single Sample Upset) | 13 | 11 | 2 |
| No. of SEU Events (with Multiple Sample Upset) | 0 | 2 | 0 |
| Total Beam On Time [s] | 1848 | 1707 | 1117 |
| TID [$kRad(Si)$] | 300 | 281 | 181 |

Table. 2. SEE test results of three ADS5272 samples with proton beam at IUCF.

For the SEU (bit flip), ADCs were characterized in no external intervention mode, i.e., a single bit or multiple bits in the data stream flips but ADCs continue to operate normally. Its impact can be examined by the measurement of cross sections. For SEFI, it was recorded when ADCs cease operation - the ADC output remains constant, requiring an external reset to bring it back to normal mode. We should notice that the SEFI here is not equal to a latch up as it doesn't need a power cycle for the ADC to recover. The ADS5272 can be reset in 200ns without a power cycle.

The first two samples were tested for cross section measurements without any special test conditions. The third sample was tested to evaluate the effectiveness of a mitigation strategy for SEFI and conditions modified to favor SEFI rather than detecting SEUs. A ~1Hz hardware reset was issued to clear any register that might be corrupted by SEU outside of the data stream. Test results are listed in Table 2.

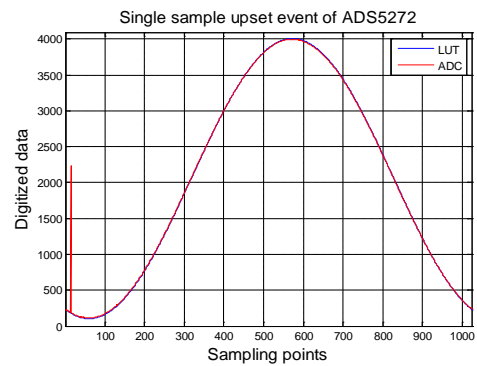

Fig. 12. Single sample upset event of ADC. Just a single sample off, but otherwise the ADC is working normally

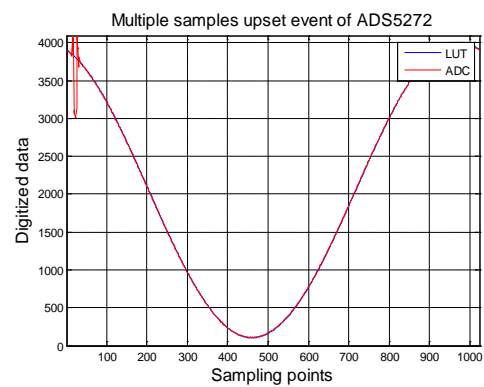

Fig. 13. Multiple samples upset event of ADC, after that ADC recovered to normal working mode

The total SEU cross section is $\sigma_{SEU} = (4.0 \pm 0.7) \times 10^{-12} cm^2$. Although with poor statistics we observe that the upset probability is independent of the bit position in one ADC word (12bits). Therefore we specified SEU cross section in units of area per bit $\sigma_{SEU}/bit = (3.3 \pm 0.6) \times 10^{-13} cm^2$. Graphical examples of SEU and SEFI are showed in Fig. 12, Fig. 13 and Fig. 14.

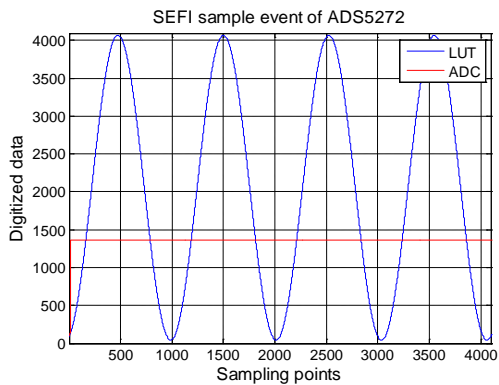

Fig. 14. SEFI event of ADC, output is a constant value. The ADC recovered to normal working mode after hardware reset

## 5 Conclusion

In this paper, an evaluation system of ADS5272 has been established and its radiation performance has been characterized. From the irradiation test results, the ADS5272 performs very well up to $300 kRad\ (Si)$ TID and up to $5 \times 10^{12} p/cm^2$ fluence without significant performance degradation. These characteristics meet the radiation tolerance criteria of the COTS component for the LAr calorimeter front end electronics [1]. Therefore, the ADS5272 has been identified as a good candidate to be used in the future LAr calorimeter electronics upgrade, and a demonstrator LTDB is now being designed with this ADC. Valuable experience and information will be obtained from this demonstrator system after installing and running it at the high-luminosity of $\mathcal{L} = 10^{34} cm^{-2} s^{-1}$ on the ATLAS LAr calorimeter.

## Acknowledgment

The authors would like to deeply thank James Kierstead, Francesco Lanni, Sergio Rescia, Hao Xu and Helio Takai of Brookhaven National Laboratory, Thomas Schwarz of University of Michigan for their help regarding this work over the years.

## References

1. Aleksa M, Cleland W, Enari Y, Fincke-Keeler M, Hervas L, Lanni F, Majewski S, Marino C, Wingerter-Seez I, et al. CERN-LHCC-2013-017; ATLAS-TDR-022
2. J. R. Schwank, P. E. Dodd, M. R. Shaneyfelt, et al. IEEE Trans. Nucl. Sci., vol. 51, no. 6, pp. 3692–3700, Dec. 2004
3. P.Koga, P. Yu, J. George, et al. Radiation Effects Data Workshop, 2008 IEEE pp. 69–75, July. 2008
4. P.Koga, P. Yu, K. Crawford, et al. IEEE Trans. Nucl. Sci., vol. 49, no. 6, pp. 3135–3141, Dec. 2002
5. Boley,W.R. Radiation Effects Data Workshop, 2008 IEEE pp. 142–147, July. 2008
6. ADS5272 8-Channel 65 MSPS Analog to Digital Converter. Texas Instruments, Dallas, Texas, Jan. 2009
7. Kai Chen, Hucheng Chen, Xueye Hu, et al. 2013 IEEE Nuclear Science Symposium and Medical Imaging Conference, Seoul, Korea, Kor, 26 Oct – 2 Nov 2013
8. ML605 Hardware User Guide, UG534 (v1.8), Oct. 2012
9. Virtex-6 FPGA Configurable Logic Block User Guide, UG364 (v1.2), Feb. 2012